\documentclass[prl,aps,showpacs,preprintnumbers,amsmath,amssymb,nofootinbib,showkeys,twocolumn,floatfix]{revtex4-2}

 \usepackage[caption=false]{subfig}
\usepackage{capt-of}
\usepackage{hhline}
\pdfoutput=1
\usepackage{xcolor}
\usepackage{hyperref}
\usepackage{natbib}
\usepackage{subfiles}
\usepackage{aas_macros}
\usepackage{amssymb}

\usepackage{graphicx}
\usepackage{dcolumn}
\usepackage{bm}
\usepackage{cases}

\hypersetup{
     colorlinks   = true,
     citecolor    = magenta,
     urlcolor     = magenta,
     linkcolor    = magenta
}

\newcommand{\sect}[1]{\noindent\textbf{\textit{#1--}}}
\newcommand{\JADEStwelve}{JADES-GS-z12-0~}
\newcommand{\JADESeleven}{JADES-GS-z11-0~}
\newcommand{\JADESten}{JADES-GS-z10-0~}
\newcommand{\JADESzthirteen}{JADES-GS-z13-0~}

\newcommand{\unit}[1]{\mathrm{#1}}
\newcommand{\GeV}{\unit{GeV}}

\newcommand{\percc}{\unit{cm}^{-3}}

\newcommand{\Msun}{M_{\odot}}

\newcommand{\Lyalpha}{Ly-$\alpha$ }
\newcommand{\microm}[1]{\unit{#1}{\mu m}}

\begin{document}
\preprint{NORDITA-2023-025}
\preprint{UTWI-13-2023}

\title{Supermassive Dark Star candidates seen by JWST?} 

\author{Cosmin Ilie}
\email[E-mail at: ]{cilie@colgate.edu}
\author{Jillian Paulin}
\affiliation{ Department of Physics and Astronomy, Colgate University\\
13 Oak Dr., Hamilton, NY 13346, U.S.A.
}%
\author{Katherine Freese}
\affiliation{Weinberg Institute for Theoretical Physics, Texas Center for Cosmology and Astroparticle Physics, \\
Department of Physics, University of Texas, Austin, TX 78712, USA;}
\affiliation{Department of Physics, Stockholm University, Stockholm, Sweden;}
\affiliation{Nordic Institute for Theoretical Physics (NORDITA), Stockholm, Sweden}
\begin{abstract}
The first generation of stars in the Universe is yet to be observed. There are two leading theories for those objects that mark the beginning of the cosmic dawn: hydrogen burning Population~III stars and Dark Stars, made of hydrogen and helium but powered by Dark Matter heating. The latter can grow to become supermassive ($M_\star\sim 10^6\Msun$) and extremely bright ($L\sim 10^9L_\odot$). We show that each of the following three objects: \JADESzthirteen, \JADEStwelve, and \JADESeleven (at redshifts $z\in[11,14]$) are consistent with a Supermassive Dark Star interpretation, thus identifying, for the first time,  Dark Star candidates.  
\end{abstract}
\maketitle

\sect{Introduction}The James Webb Space Telescope (JWST) is poised to revolutionize  our understanding of the formation and properties of first luminous objects in the universe. Since beginning to take data, JWST has discovered a surprising number of extremely bright high redshift galaxy candidates ~\citep[e.g.][]{GLASSz13,Maisies:2022,z16.CEERS93316:2022,z17.Schrodinger:2022,Labbe:2022}, 
which are difficult to reconcile with expectations from numerical simulations of the universe in the canonical $\Lambda$CDM scenario.
 In this paper we show that some of the JWST high redshift galaxy candidates could instead be Dark Stars (DSs), early stars made almost entirely of hydrogen and helium but powered by dark matter (DM) heating rather than by fusion. Dark stars provide a good  match to JWST data, both in terms of their spectra (good fits to JWST photometry) and in that JWST (given its angular resolution) cannot rule out a point source interpretation of many of these candidates.

Prior to JWST we had very limited data on the cosmic dawn era, i.e. the period when the first stars and galaxies form. As such, numerical simulations were the primary tool to describe the properties of the first stars~\citep[e.g.][]{Abel:2001,Barkana:2000,Bromm:2003,Yoshida:2006,OShea:2007,Yoshida:2008,Bromm:2009}, and galaxies~\citep[e.g.][]{Loeb:2010,Bromm:2011,Gnedin:2016,Dayal:2018,Yung:2019,Behroozi:2020} in the Universe. In the standard picture of the first (Population~III a.k.a Pop~III) stars, they formed roughly 100-400 Myrs after the Big Bang ($z \sim 20-10$) as a consequence of the gravitational collapse of pristine, zero metallicity, molecular hydrogen  clouds at the center of $10^6-10^8 M_\odot$ minihaloes. Pop~III stars grow via accretion, reaching masses of (at most) $10^3\Msun$~\cite[e.g.][]{Hirano:2014}, and populate the first galaxies.  We previously proposed, however, a different type of first star ~\citep{Spolyar:2008dark,Freese:2008ds,Spolyar:2009}:  Dark Stars, early stars powered by dark matter heating rather than by fusion.

Some of the JWST high redshift galaxy candidates could instead be Dark Stars, which are made almost entirely of hydrogen and helium with less than 0.1\% of the mass in the form of dark matter.  Since they remain cool (without a central hot core), there is no fusion inside them; instead, DM annihilations happen throughout their volume.  Dark stars are giant, puffy ($\sim$ 10 AU) and cool (surface temperatures $\sim10,000$ K) objects. We follow the evolution of dark stars, in thermal and hydrostatic equilibrium, from their inception at $\sim 1 M_\odot$ as they accrete mass from their surroundings to become Supermassive Dark Stars (SMDSs), some even reaching masses $> 10^6 M_\odot$ and luminosities $> 10^{10} M_\odot$, making them visible to JWST~\cite{Freese:2010smds,Ilie:2012}. Once the dark matter runs out and the SMDS dies, it collapses to a black hole; thus dark stars may provide seeds for the supermassive black holes observed throughout the Universe and at early times. 

As of this writing, out of the hundreds of potentially high-$z$ objects discovered by JWST, only $\sim$ 10 have been confirmed spectroscopically via the identification of the Lyman-break feature in their SEDs. In particular, the JWST Advanced Deep Extragalactic Survey (JADES) has discovered four spectroscopically confirmed Lyman break objects:
\JADESzthirteen, \JADEStwelve, \JADEStwelve, and \JADESten~\cite{JADES:2022a,JADES:2022b}. 
In this paper we will show that three of these four JADES high-$z$ objects are consistent with Supermassive Dark Stars.
We find that, with the exception of \JADESten, the photometry of those objects can be modeled by SMDSs Spectra.

For a more definitive answer, higher quality spectroscopy of the objects will be required. Specifically, as we will show below, the appropriate Helium lines could be smoking guns for dark stars, and could distinguish between, for instance, a galaxy made of Pop~III stars and a single SMDS. A Helium-II absorption feature at $1640~\unit{\AA}$ would be characteristic of a hot SMDS~\cite{Ilie:2012}.  An emission line at the same wavelength would be characteristic of a Pop~III galaxy. For all types of SMDSs, the Balmer absorption lines, at rest frame wavelengths $\lambda\gtrsim \microm{0.35}$ can be used to differentiate them from early galaxies, which will typically exhibit emission lines at the same wavelengths~\citep{Zhang:2022}.  

\sect{Dark Stars}As the molecular clouds of hydrogen collapse inside early minihaloes in the process of star formation, the large reservoir of dark matter at the centers of the minihaloes can play an important role. If the DM particles are their own antiparticles, then their annihilation provides a heat source that stops the collapse of the clouds and in fact produces a different type of star, a Dark Star, in thermal and hydrostatic equilibrium.  We wish to  emphasize that Dark Stars are made almost entirely of ordinary matter (hydrogen and helium) but powered by DM, even though the DM only constitutes less than 0.1\% of the DS.  We considered two types of DM particles:  Weakly Interacting Dark Matter (WIMPs, in most of our papers) and Self Interacting Dark Matter (SIDM). 
The energy production per unit volume provided by the annihilation of two DM particles is given by:  
\begin{equation}\label{eq:heat}
    Q = m_\chi n_\chi^2 \langle \sigma v\rangle = \langle \sigma v\rangle \rho_\chi^2 / m_\chi,
\end{equation}
 where $m_\chi \sim 1 {\rm GeV} - 10 {\rm TeV}$ is the DM mass, $n_\chi$ is the DM number density, $\rho_\chi$ is the DM energy density.  We have used the fact that
the DM mass is converted to energy in the annihilation, 
and we took the standard annihilation cross section (the value that produces the correct DM abundance in the Universe today): $\langle \sigma v\rangle = 3 \times 10^{-26} {\rm cm}^3/{\rm s} \, .$ We note that cross sections several orders of magnitude smaller or larger would work equally well; by considering a variety of WIMP masses we can see from Eq.(\ref{eq:heat}) that this is equivalent to considering a variety of cross sections. Three key ingredients are required for the formation of DSs: (1) sufficient DM density, (2) DM annihilation products become trapped inside the star, and (3) the DM heating rate beats the cooling rate of the collapsing cloud. In our previous work we showed that all three criteria can be easily met.  

The  criterion of high DM density can be met in two ways. First, as the hydrogen cloud collapses, it dominates the potential well and pulls in more DM with it. This phenomenon can be well described by adiabatic contraction (AC).\footnote{We have confirmed the results of the simple~\citeauthor{Blumenthal:1985}~\cite{Blumenthal:1985} AC approach via more sophisticated analyses~\citep{Freese:2008dmdens}.}  Since many DM particles are on chaotic or box orbits, the central DM density can be replenished and kept high for millions (to billions) of years.  Secondly, once the DM power is depleted, the star starts to collapse, and in the process reaches a high enough density that it is able to capture further DM particles via elastic scattering of the DM with the atoms in the star.  We will consider both extended AC and capture in this paper.

Once a DS forms of $\sim 1 M_\odot$, we have studied its evolution with two different types of stellar codes:  one of which assumes that the DS can be approximated as a polytrope, and the MESA stellar evolution code~\citep{Paxton2011}.  In both cases, we find essentially the same results~\cite{Rindler-Daller:2015SMDS}. Because the DS are puffy and cool (surface temperatures $\sim 10^4$K), they
are able to accrete the material around them and become very massive (there is not enough ionizing radiation to prevent accretion).  We find the equilibrium structure for the stars of a given mass, and then build up the stars one solar mass at a time, always in equilibrium, and find that some of them can become Supermassive Dark Stars (SMDS) that are incredibly massive ($>10^6 M_\odot$) and bright ($> 10^9 L_\odot$), and the heaviest ones should be visible in JWST. Because they are simultaneously bright, they may look different from competing objects. 

\sect{Method}We describe below our method to look for Dark Star candidates in JWST data. High redshift luminous objects are typically discovered as photometric dropouts, in deep field surveys\footnote{In other words, the objects are detected in one redshfit bin but not in a neighboring one, due to absorption by Ly$-\alpha$ along the line of sight. The wavelength of the observed dropoff in flux (redshifted from the time of the emission of the light) is then used to identify the redshift of the object.}. Our focus in this paper is dropout candidates identified already in the literature that satisfy the following two criteria: i) a Lyman break has been spectroscopically identified, such that the objects are definitively at high ($z\gtrsim10$) redshift\footnote{Thus we avoid potential low redshift interlopers as another possible interpretation of the photometric data. Dropping condition i) would allow us to find more SMDS candidates, and we plan to do so in a future publication.}, and  ii) the objects are unresolved, or marginally resolved, so that they can be consistent with an explanation in terms of a point object.
Any object that satisfies both criteria can be consistent with the hypothesis that it is a Dark Star.

We have found three candidates that match both criteria.
Specifically, of the four spectroscopically confirmed Lyman break objects: \JADESzthirteen, \JADEStwelve, \JADEStwelve, and \JADESten~\cite{JADES:2022a,JADES:2022b}, all are consistent with possibly being point objects, and three have photometry that
can be modeled by SMDS spectra (with the exception of \JADESten).

{\it{Resolved vs unresolved objects:}} SMDSs would be point objects in JWST data whereas galaxies are larger and hence may be resolved. The angular resolution of  JWST is approximately $\theta_{res}\sim10^{-6}$ radians~\citep{JWST:2006}. At $z\simeq 10$ a SMDS (with radius $R\sim 10$ AU) will have an unlensed angular size of $\sim 10^{-13}$ radians, well below the angular resolution of any imaginable telescope. Even if strongly lensed, SMDSs will still be below the resolution limit of JWST, i.e. look like point sources. Some of the  galaxy candidates identified in the JWST data are barely resolved, or unresolved, i.e. consistent with point sources.  

The four JADES objects we are considering are all consistent with a point source interpretation.
For two of them, \JADESeleven, and \JADEStwelve, the authors argue that they are resolved, under the assumption of an interpretation in terms of galaxies. However, their estimated effective sizes ($\sim 0.02''$ and $\sim 0.04''$, approximately the size of one NIRCam pixel) are about one order of magnitude below the resolution limit of JWST NIRCam ($\sim 0.1''$). We therefore consider here the possibility they are unresolved, point sources.   While these objects are too dim to see Airy patterns that would definitively identify them as point objects, it would be interesting if future observations were able to make this distinction.

{\it Dark Star Spectra:} Our next task is to show that the spectra of three of the four objects are well fit by Dark Star spectra.
We note here the spectra for those four objects, obtained in~\cite{JADES:2022b}, do not yet confidently identify any spectral lines, as they are too noisy (S/N$\sim 2$). Followup spectroscopy is required in order to determine the presence of emission/absorption features. For this reason we restrict our discussion in this paper to comparing SMDSs to photometric data.

The spectra of SMDSs were obtained using the \textsc{TLUSTY}~\citep{Hubeney:1988} synthetic stellar atmospheres code. This code accounts for not only the blackbody radiation from the photosphere of the DS, but also for absorption or emission features (lines and breaks) arising from the gas in the atmosphere of the star. For each SMDSs formation mechanism we initially generated TLUSTY SEDs on a coarse stellar mass grid: $\sim 10^4, 10^5, 10^6, 10^7$ (in units of $\Msun$)~\cite{Ilie:2012}. If needed, in order to get an optimal fit, we  further refined the grid to include midpoints. In this study we take DM mass $m_\chi$ = 100 GeV in obtaining SMDS spectra. However, in future works we plan to consider a variety of other DM particle masses, some of which would  increase the likelihood of finding good photometric fits to JWST data. Below we discuss the main features of the SMDS SEDs (see also Fig.~\ref{fig:spectra} in Supplementary Material).
For the SMDSs formed via DM capture we find that the  SEDs are nearly independent of the stellar mass, since their temperatures are roughly constant. They also have a much steeper slope of the (UV) continuum when compared to the cooler SMDS formed via AC.

An important tool in the future to differentiate SMDSs from early galaxies will be the
the He~II $\lambda$1640 line at $\microm{0.1640}$ (restframe), present for all SMDS formed via DM capture and for the $M_\star\gtrsim 5\times 10^5\Msun$ SMDS formed via AC. By contrast with the He~II $\lambda$1640 line from SMDSs, galaxies would typically exhibit a nebular emission line at the same wavelength. For the cooler SMDSs (formed via AC) the strong lines in the H Balmer series, redward of the Balmer break  at $\microm{0.36}$ can be SMDS smoking guns, as Pop~III/II galaxies typically exhibit strong nebular emission lines at same wavelengths.

In order to convert the rest frame SEDs of each SDMS considered into observable quantities we first redshift them:  
\begin{equation*}\
    F_{\nu}(\lambda_{obs} ; M_\star; z_{emi})=\frac{(1+z_{emi}) 4 \pi R_{*}^2 F_{\nu}\left(\lambda_{emi};M_\star\right)}{4 \pi D_{\mathrm{L}}^{2}(z_{emi})}, 
\end{equation*}
where $\lambda_{obs}=(1+z_{emi})\lambda_{emi},$ and $\lambda_{emi}$ represents the observed and emitted wavelength, $R_{*}$ is the radius of SMDSs, $M_\star$ is its radius, and $D_{\mathrm{L}}$ is the luminosity distance. $ F_{\nu}\left(\lambda_{emi}; M_\star\right)$ represents the rest frame flux density (plotted in Fig.~\ref{fig:spectra}) and and $F_{\nu}(\lambda_{obs} ; z_{emi})$ is the redshifted flux density. We account for the Gunn-Peterson trough~\citep{Gunn-Peterson:1965} by suppressing all the flux short-ward of the redshifted \Lyalpha line to zero, since all of the candidates we will be looking for are at $z_{emi}\gtrsim 10$.

{\it Comparing SMDS to JWST data:} In each JWST band (labeled here by the letter b) we find the average expected redshifted flux due to a SMDS:
\begin{equation*}
    \tilde{F}_{\nu;b}(M_\star;z_{emi})=\frac{\int_{\lambda_{min}}^{\lambda_{max}} T(\lambda_{obs}) F_{\nu}(\lambda_{obs} ; M_\star; z_{emi}) \frac{\mathrm{d} \lambda_{obs}}{\lambda_{obs}}}{\int_{\lambda_{min}}^{\lambda_{max}} T(\lambda_{obs}) \frac{\mathrm{d} \lambda_{obs}}{\lambda_{obs}}}
\end{equation*}
where $\lambda_{obs}$ is the observed wavelength, $T(\lambda_{obs})$ is the throughput curve for the photometric band (filter) in question, and $\lambda_{min;max}$ represent the two wavelengths defining the band-pass filter denoted here by the letter $b$.

{\it Best fit Dark Star Models:} We then compare the predicted SMDS photometry on our stellar mass and formation mechanism grids to JWST data in all bands available for each candidate analysed. Some of the JWST objects may be gravitationally lensed, which will enhance the flux by magnification factor $\mu$.
Therefore, we optimize the fit between model ($F_{\nu;b}$) and data ($f_{\nu;b}$) with regard to two main parameters: the redshift $z_{emi}$ and  $\mu$. Using a $\chi ^2$ analysis:
\begin{equation}\label{eq:chi2}
\chi^2=\sum_b\frac{(f_{\nu;b}-\mu\times\tilde{F}_{\nu;b}(M_\star; z_{emi})^2}{\sigma^2(f_{\nu;b})+\sigma^2_{sys}(b)}
\end{equation}
we can determine which combinations of $z_{emi}$ and $\mu$ align most closely with the data. In Eq.~\ref{eq:chi2} we sum over all bands for which photometric data is available, and $\sigma(f_{\nu;b})$ represents the statistical flux error in a given band, whereas $\sigma_{sys}(b)$ represents the systematic error term, that accounts for zero point uncertainties, or imperfect aperture corrections. 

For each SMDS formation mechanism considered (AC or DM capture) we select the  stellar mass in our TLUSTY grid for SMDS SEDs that will lead to a value of the $\mu$ parameter closest to 1, as the objects analysed in this work are typically assumed to be unlensed. However, in reality, they could be lensed by $\mu$ as large as $\mathcal{O}(10)$. While there are no {\it known} foreground clusters or galaxies that could act as lenses for the objects considered here (\JADESeleven, \JADEStwelve and \JADESzthirteen), the fact that the objects are unresolved leaves open the possibility that they are indeed strongly lensed, with the resulting shear undetectable with current resolution. On the other hand, most lines-of-sight in the universe will have $\mu<1$
(with photons being pulled away from the line-of-sight towards nearby overdense regions) \citep{Holz:1998,Wang:2002}. N-body simulations reveal that the probability of a given $\mu$ is a redshift dependent function ($P(\mu;z)$) for which the peak is moving towards lower values as $z_{emi}$ is increased. For instance, at $z=3.2$ the peak of $P(\mu)$ is around $\mu\simeq 0.9$~\citep{Wang:2002}, with values as low as $\mu\simeq 0.7$ not ruled out. By $z\sim 10$  even lower values of $\mu$ will not only possible, but likely. 

There is a degeneracy between $\mu$ and $M_\star$ with regards to the SMDSs models. For SMDSs formed via capture, this degeneracy is almost one to one, since the rest frame flux is largely independent of $M_\star$ (see Fig.~\ref{fig:spectra}). As such, the redshifted fluxes of SMDSs formed via capture, at any given wavelength, scale linearly with either $M_\star$ or $\mu$. The degeneracy is still present for the SMDSs formed via AC, albeit less trivial, since now there is a stellar mass dependence of the rest frame fluxes. In summary, scanning over $\mu$ is a proxy for scanning over $M_\star$ while keeping $\mu$ fixed. In other words, for any optimal fit we find for the three parameters: $\mu_{fit}$, $z_{fit}$, and $M_\star$ there will be an equally good fit with $\mu\approx1$, $z_{fit}$, and a somewhat larger or smaller value of $M_\star$, depending if $\mu_{fit}$ is smaller or larger than one.

\sect{Results} Three of the four JADES objects considered, \JADESzthirteen,  \JADEStwelve, and \JADESeleven, have photometry consistent at a minimum 95\% confidence level (CL) with a Dark Star interpretation, as shown in Figure~\ref{fig:Heatmaps}. In the upper panels  we have plotted the $\chi^2$ for the match in the $\mu$ vs $z$ plane, where all points shown in the colored region fit the data at the 95\% CL, or better for each of the three objects as labeled. 
\begin{figure*}[!htp]
\centering
\includegraphics[width=.32\textwidth]{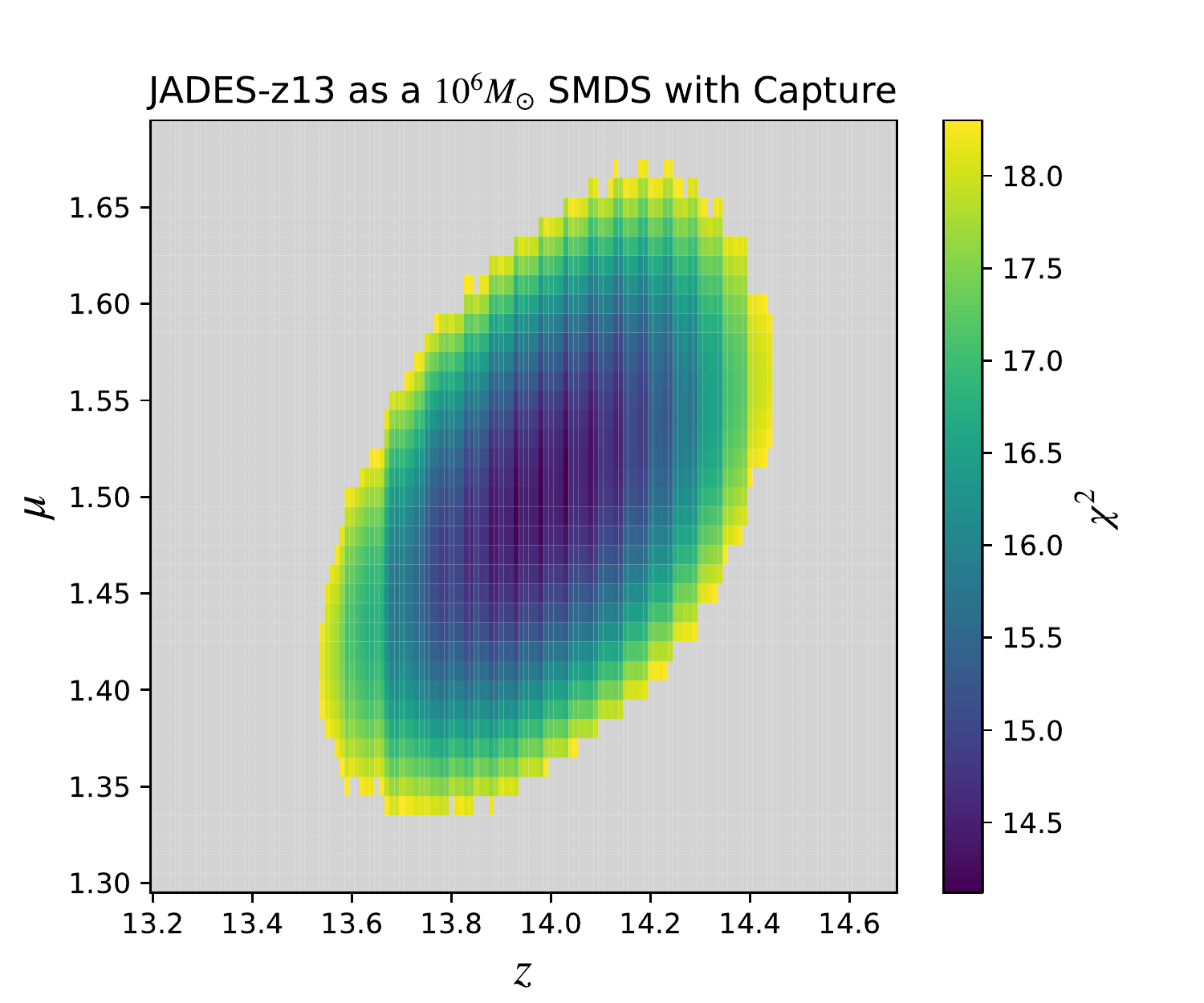}
\includegraphics[width=.32\textwidth]{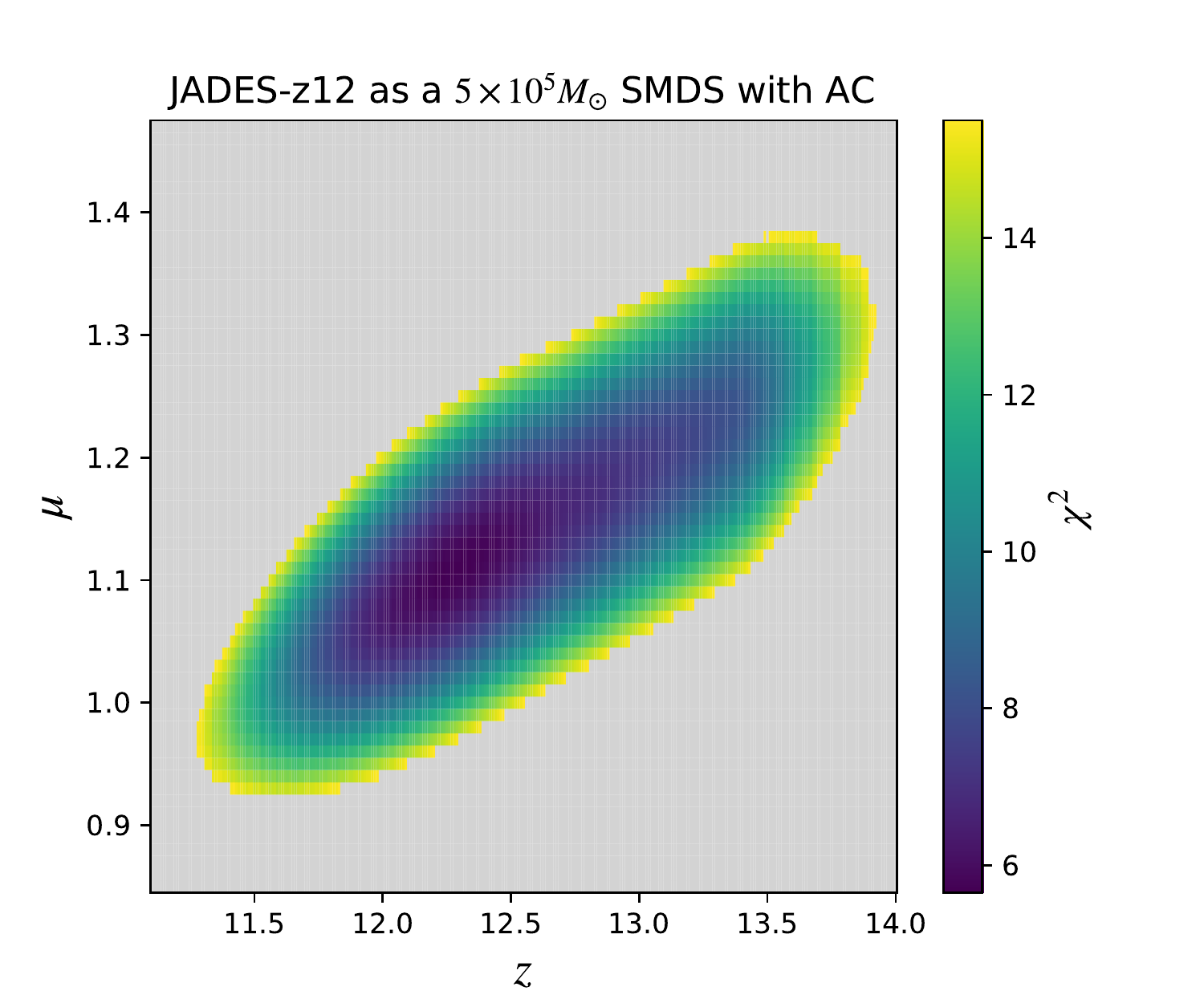}
\includegraphics[width=.32\textwidth]{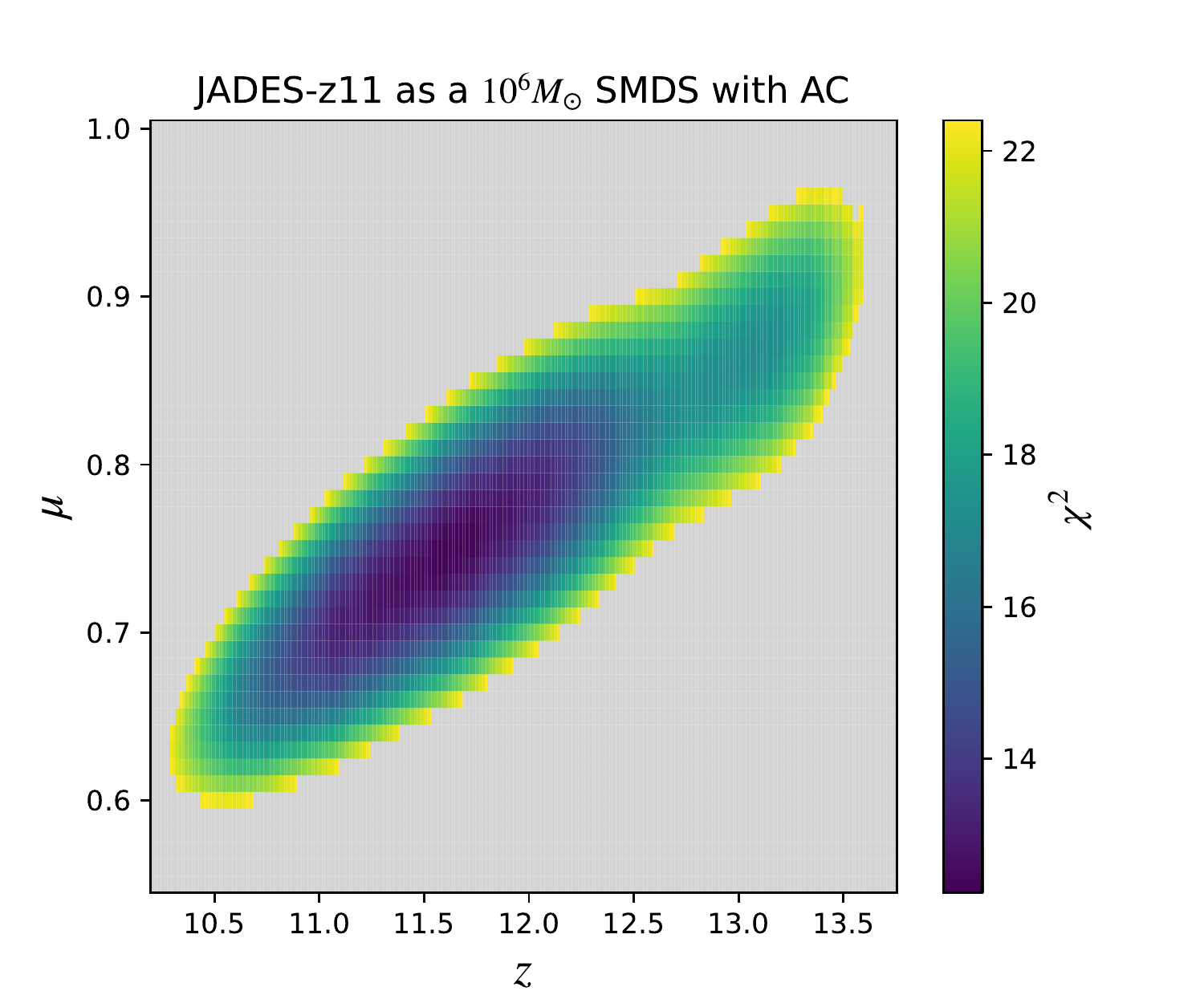}\\
\includegraphics[width=.32\textwidth]{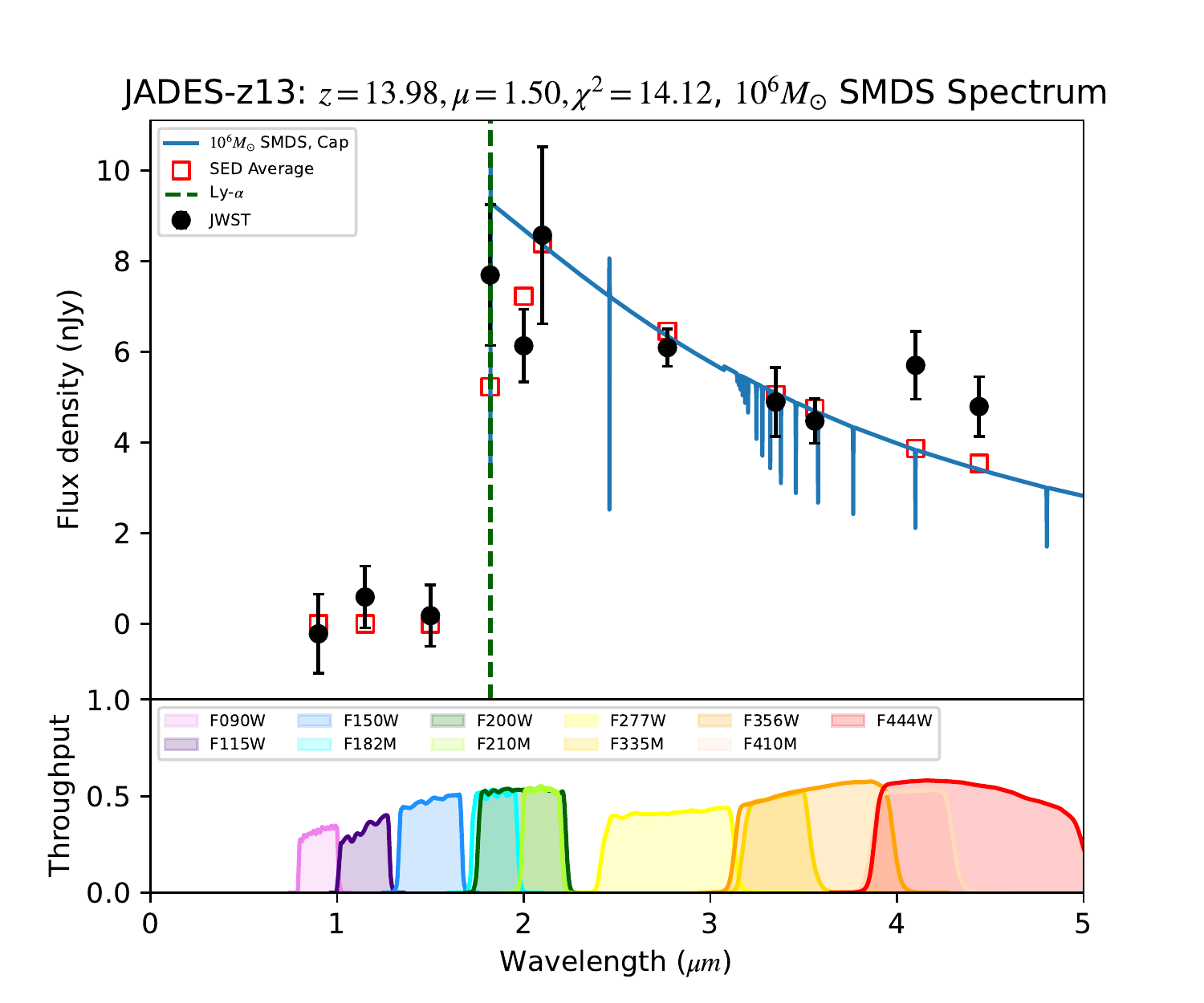}
\includegraphics[width=.32\textwidth]{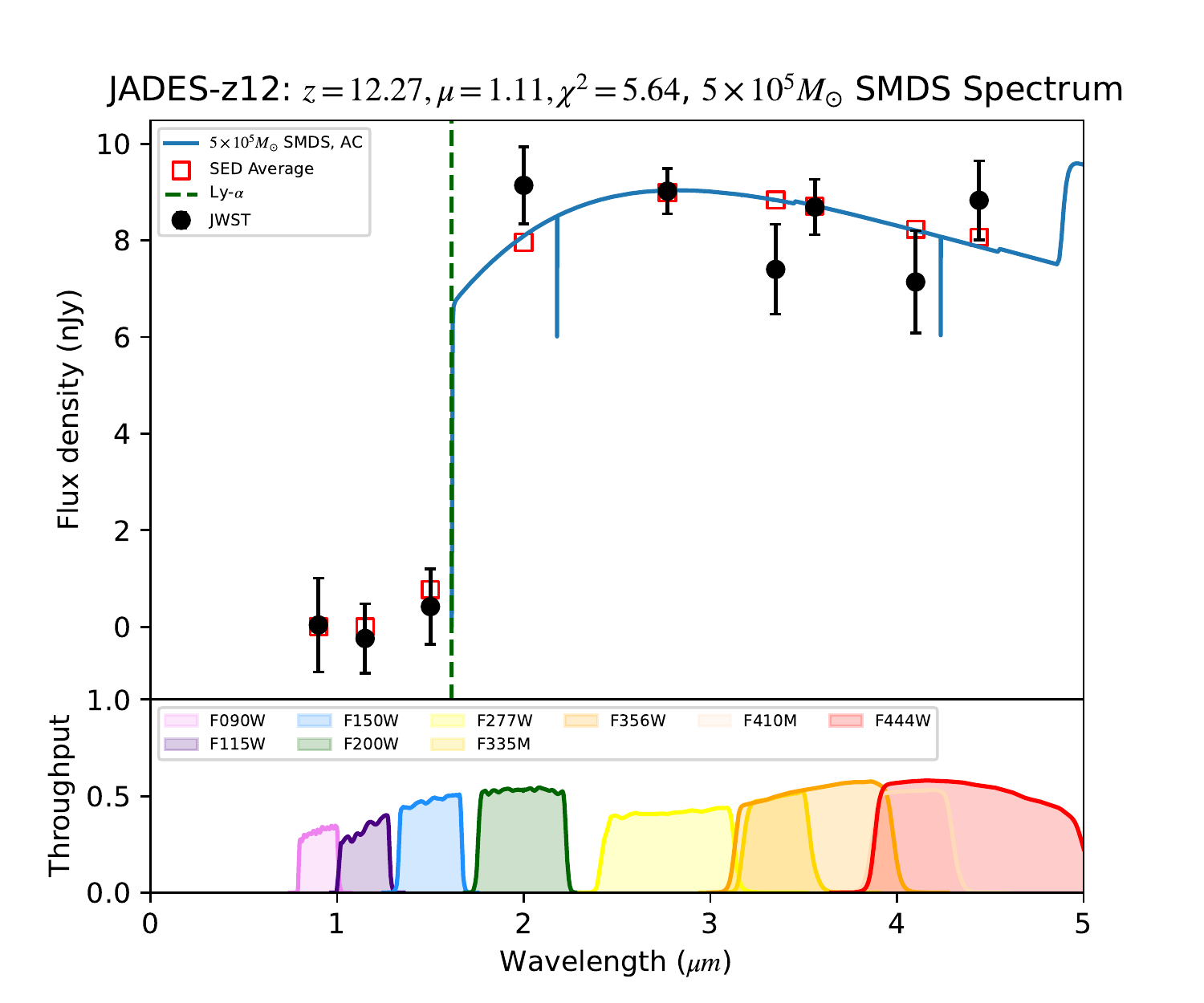}
\includegraphics[width=.32\textwidth]{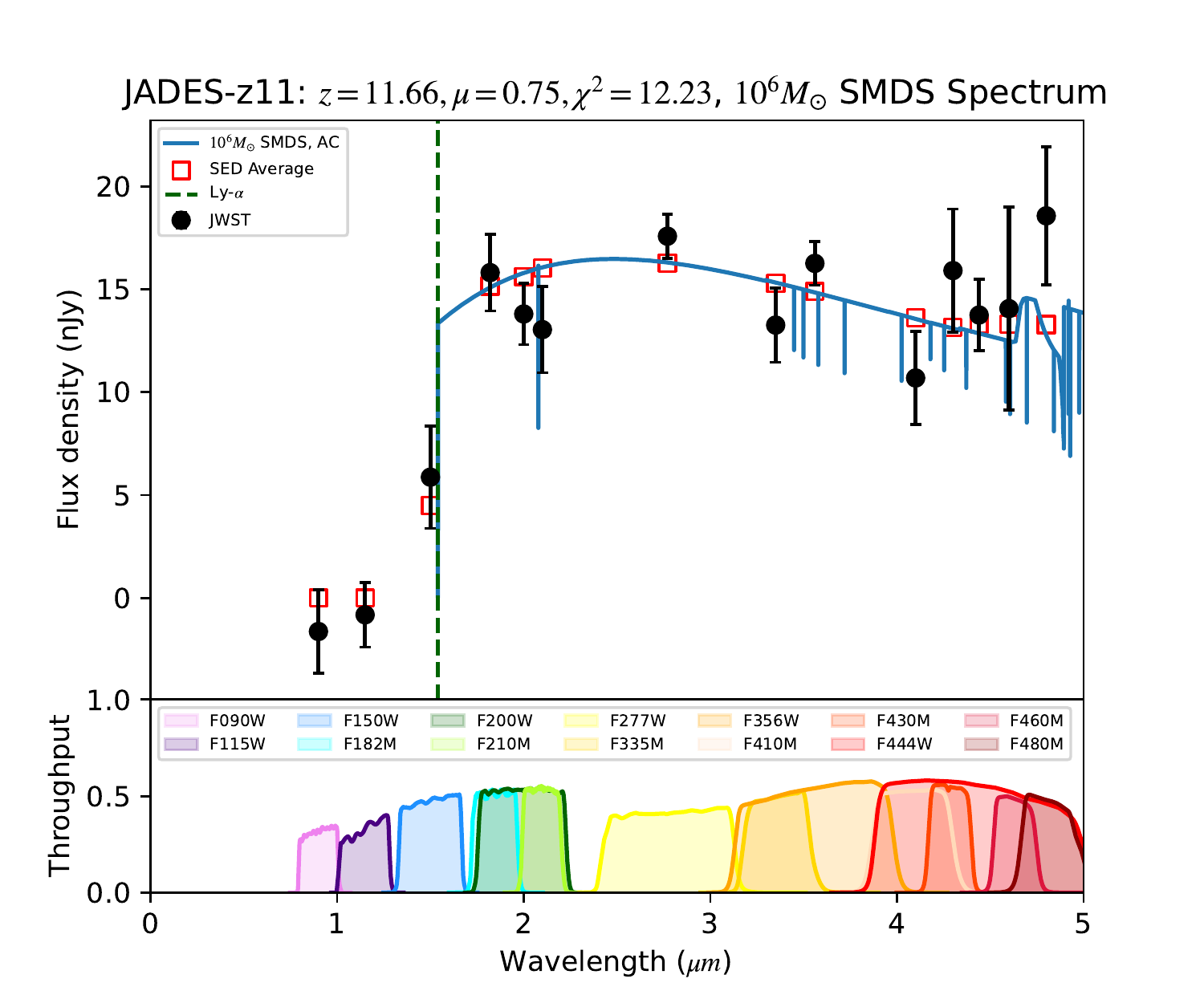}
\caption{ (Top Row) Optimal fit regions in the $z$ vs   $\mu$ (magnification) parameter space for Supermassive Dark Star fits to  \JADESeleven, \JADEStwelve, and \JADESzthirteen  photometric data. The heatmap is color coded according to the value of the $\chi^2$, and is cut off (grayed out) at the critical value corresponding to 95\% CL. In addition to labeling the object, the title in each panel includes the the mass and formation mechanism for the SMDSs model considered. (Bottom Row) For each case we plot our best fit SEDs against the photometric data of~\cite{JADES:2022a} in each band. Titles include values of relevant parameters and $\chi^2$. Each band is visually represented by its throughput curves, color coded and plotted at the bottom of the SED plots.}
\label{fig:Heatmaps}
\end{figure*}

{\it {\bf Best Fit Parameters:}} For \JADESzthirteen (left panels) we find that a $10^6\Msun$ SMDS formed via DM capture, at $z_{photo}\simeq 13.98$, boosted by a gravitational lensing factor of $\mu\simeq 1.5$ leads to our best fit, with a $\chi^2\simeq 14.12$. For this object there are photometric data in eleven JWST bands, so that 95\% C.L. corresponds to the critical value $\chi^2\simeq 18.3$. For \JADEStwelve our best fit is a $5\times 10^5\Msun$ SMDS formed via AC, at $z_{photo}\simeq 12.27$, lensed by a factor of $\mu\simeq 1.11$, with a $\chi^2\simeq 5.64$. For this object there is photometric data in 9 bands, corresponding to $\chi^2_{crit}\simeq15.51$. Of all the objects considered, this is by far our best fit. Lastly, for \JADESeleven our best fit is a $10^6\Msun$ SMDS formed via AC, at $z_{photo}\simeq 11.66$, de-lensed by a factor of $\mu\simeq 0.75$,\footnote{If we refined our SMDS mass grid we would be able to find an equally good or better fit with slightly lower $M_\star$ and $\mu\gtrsim 1$. Yet at high redshifts typical objects have a $\mu<1$~(see Fig.~4 of~\cite{Wang:2002}).}  and with a $\chi^2\simeq 12.23$
(and $\chi_{crit}\simeq 22.36$). 

We note that in this paper we have only considered a coarse grid of stellar masses. In the future, we should be able to find even better
fits to the data by considering a finer grid of SMDS masses. The aim of the current paper is to point out that SMDS SEDs are excellent fits to JWST photometric data for Lyman break objects, not to find the best possible Dark Star fit to a given set of data. 

{\it Consistency with $z_{spec}$:} We now compare our best fit redshifts $z_{photo}$ obtained by fits of SMDS SEDs to JWST photometry, to the spectroscoptic redshifts
$z_{spec}$ estimated by~\cite{JADES:2022b}, for each of the three candidate objects, in turn.

{\noindent\it\JADESzthirteen:} For this object $z_{spec}\simeq 13.2$, which is outside of our 95\% CL.  (see top left panel in Fig.~\ref{fig:Heatmaps}). However, there are several reasons not do discard \JADESzthirteen as a SMDS candidate. First of all, the estimation of $z_{spec}$, given the $S/N\lesssim2$ of the spectra for \JADESzthirteen (see bottom right panel of Fig.~1 of~\cite{JADES:2022b}), has inherent uncertainties. Those will will be significantly reduced once cleaner spectra are obtained.\footnote{For instance, for GN-z11, $z_{spec}\simeq 11.1$ from HST Grism~\cite{GNz11}, whereas the exquisite spectrum obtained recently with NIRSpec by~\cite{GNz11NIRSpec:2023} leads to a more accurate $z_{spec}\simeq 10.60$} Furthermore, our $z_{photo}$ will be slightly lower if we had a finer TLUSTY SED grid, with a slightly higher $M_\star$. Lastly, we have verified that $z_{spec}\simeq 13.2$ is within the 98\% CL region, so statistically our best fit is not ruled out at the $2-\sigma$ level, even without accounting for the uncertainties explained above, that could significantly reduce this mild tension.

{\noindent\it\JADEStwelve:} In this case $z_{spec}\simeq 12.63$, consistent with our $z_{photo}\simeq 12.27$, as it falls well inside the 95\% CL (middle panel of Fig.~\ref{fig:Heatmaps}). 

{\noindent\it\JADESeleven:} For this object we find again, excellent agreement between $z_{spec}\simeq 11.58$ and our best fit $z_{photo}\simeq 11.66$. 

{\noindent\it {\bf Spectral signatures of SMDSs candidates:}} In Fig.~\ref{fig:Heatmaps} (bottom row) we plot the photometric data (solid circles with error bars corresponding to the statistical uncertainty in each measurement) from~\cite{JADES:2022a} against our SMDS models (red squares representing the average SMDSs flux in each band). The blue lines represent the redshifted (at $z_{photo}$) best fit SMDS SEDs. Followup spectroscopy with high S/N could potentially identify the features in the SMDS spectra which would differentiate them from galaxies. A smoking gun, for {\it{all three}} SMDS candidates, is the He~II $\lambda1640$ absorption line, which will either be absent in the case of galaxies without nebular emission, or become an emission line for galaxies with strong nebular emission. Below we discuss other spectral features for each candidate.  The low S/N$\sim 2$ of the spectra obtained in~\cite{JADES:2022b} does not conclusively identify any emission or absorption features, so in order to confirm their status as SMDSs we would need followup, more detailed SEDs with NIRSpec or other observatories. 

To make detailed comparison with spectra, we will also need to perform additional work on the modeling side. The TLUSTY spectra we have obtained need to also be run through CLOUDY to take into account the effect of nebula around the SMDS, in terms of both absorbing some of the light emitted by the SMDS as well as emitting light of its own. For the largest SMDS, which have accreted much of their surroundings, the nebular effects would be minimized. Consideration of the effect of nebula on the spectra is the subject of future work.

{\noindent\it\JADESzthirteen:} For the $10^6\Msun$ SMDSs via capture modelling this object we find that the spectrum includes a sequence of other He lines within the NIRSpec window, however none of them as strong as the aforementioned He~II $\lambda1640$. In general, the hottest SDMSs, such as those formed via capture, will have a very mild Balmer break, if at all, and almost no absorption or emission features in the Balmer series. However, for those objects the  He~II $\lambda1640$ is the strongest, in comparison to cooler SMDSs. 

{\noindent\it\JADEStwelve:} The $5\times 10^5\Msun$ SMDS via AC modelling this object has a nearly featureless SED, except for the absorption around $\microm{3.2}$ (redshifted). This corresponds to the He~I $\lambda3187$ line. Also, this object, with a $T_{eff}\simeq 1.7\times 10^4$~K there is a rather pronounced Blamer break, at around $\microm{4.6}$, with a jump in the flux of about 20\% that should be detectable with NIRSpec.   

{\noindent\it\JADESeleven:} The SED for the $10^6\Msun$ SMDSs via AC modelling this object has quite a few more He lines within the NiRSpec window, in comparison to the cooler $5\times 10^5\Msun$ discussed above. This is also the brighest of the three candidates, with a flux in many bands exceeding 15~nJy.

\sect{Conclusions} We have identified three SMDS candidates at $z\in[11,14]$ in the JWST data: \JADESzthirteen, \JADEStwelve, and \JADESeleven. For each of them a low redshift contaminant is excluded in view of the spectroscopic detection of the Lyman break~\cite{JADES:2022b}. Additionally, we made predictions for the SEDs of those SMDS candidates, and suggested smoking gun signatures such as the He~II $\lambda1640$ absorption line, a feature expected for all SMDSs but not for Pop~III/II galaxies. We further note that the spectra of SMDS and early galaxies differ for wavelengths above $\sim \microm{5}$, so that future observatories (beyond JWST) might be able to differentiate the two types of objects in this way. The three JADES objects are currently consistent with point objects given the limitations of the angular resolution of JWST; in the future, if Airy patterns were identified for any SMDS candidate, that might confirm its point-like nature. The confirmation of even a single one of those objects as a Dark Star (with detailed NIRSpec spectra) would mark a new era in Astronomy: the observational study of dark matter powered stars.

\sect{Acknowledgments} K.F. is grateful for support from the Jeff and Gail Kodosky Endowed Chair in Physics  at the Univ. of Texas, Austin.   K.F.  acknowledges funding from the U.S. Department of Energy, Office of Science, Office of High Energy Physics program under Award Number DE-SC0022021. 
K.F. acknowledges support by the Vetenskapsradet (Swedish Research Council) through contract No. 638- 2013-8993 and the Oskar Klein Centre for Cosmoparticle Physics at Stockholm University. We  are extremely grateful to Marcia Rieke for answering our many questions about the JADES results.  We are also grateful to Daniel Eisenstein, Steve Finkelstein, Daniel Holz, and Yun Wang for their helpful discussions. We also want to thank Saiyang Zhang for sharing his code used in~\cite{Zhang:2022} that was used to redshift TLUSTY-formatted SEDs and Luca Visinelli for sharing the TLUSTY SED for the $5\times 10^5\Msun$ SMDSs used here. Last, but certainly not least, we would like to thank our past collaborators on the subject of Dark Stars including Douglas Spolyar, Paolo Gondolo, Monica Valluri, and Tanja Rindler-Daller, who made this paper possible. 
\pagebreak[4]
\vspace{10cm}

\newpage
\maketitle
\onecolumngrid
\newpage
\begin{center}
\textbf{\large Supermassive Dark Star candidates seen by JWST?}

\vspace{0.05in}
{ \it \large Supplementary Material}\\ 
\vspace{0.05in}
{Cosmin Ilie, Jillian Paulin, and Katherine Freese}
\end{center}
\onecolumngrid
\setcounter{equation}{0}
\setcounter{figure}{0}
\setcounter{section}{0}
\setcounter{table}{0}
\setcounter{page}{1}
\makeatletter
\renewcommand{\theequation}{S\arabic{equation}}
\renewcommand{\thefigure}{S\arabic{figure}}
\renewcommand{\thetable}{S\arabic{table}}

The Supplementary Material contains additional details relevant to our work presented in the main text. Additionally, for the reader's convenience, it presents enlarged versions of our plots included in the main text, which, due to page limits, were combined into the six panel Fig.~\ref{fig:Heatmaps}. This Supplementary Material is divided into two sections: Dark Star Spectra and HR Diagram, and SMDS Candidates.

\section{Dark Star Spectra and HR Diagram}\label{sec:DSSEDs}

In Table~\ref{tab:SMDSpara} we list the relevant parameters for SMDSs on our stellar mass and formation mechanism grid that were then passed to TLUSTY to obtain relevant SEDs. 

\begin{table}[h!]
\centering
\begin{tabular}{lcccc}
 \hline
 Formation Mechanism &$M_{*}$  & $L_{*}$ & $R_{*}$  & $T_{\text {eff }}$ \\ &$\left(M_{\odot}\right)$ & $\left(10^{6} L_{\odot}\right)$ & $(\mathrm{AU})$ & $\left(10^{3} \mathrm{~K}\right)$  \\ [0.5ex] 
 \hline\hline
Extended AC & $2.04\times 10^{4}$ & 407 & 31 & 10 \\
Extended AC & $10^{5}$ & $2.42\times 10^3$ & 39 & 14  \\
Extended AC & $5\times10^{5}$ & $7.21\times10^{3}$ & 46 & 17 \\
Extended AC & $10^{6}$ & $2.01 \times 10^{4}$ & 61 & 19 \\
Capture & $4.1 \times 10^{4}$ & 774 & 1.8 & 49  \\
 Capture & $10^{5}$ & $1.75 \times 10^{3}$ & 2.7 & 51 \\
 Capture & $10^{6}$ & $2.03 \times 10^{4}$ & 8.5 & 51 \\ 
 \hline
 \end{tabular}
 \caption{Parameters of SMDSs of various masses if formed via Extended AC or DM Capture. The values listed here are adopted from~\cite{Freese:2010smds} (see Tables~3 and 4 there). We assume both types of SMDSs are powered by annihilations of 100~GeV WIMPs and formed in $10^8 M_{\odot}$ DM halos at redshift $z_{form}=15$ and grow via accretion, at a rate of $\dot{M}=10^{-1} M_{\odot} \mathrm{yr}^{-1}$. For the case of a SMDS formed via DM capture, we further assume that the product between the ambient DM density and the DM-proton scattering cross section is: $\rho_\chi\sigma=10^{13}~\GeV\percc\times 10^{-39}\unit{cm}^2$}\label{tab:SMDSpara}
\end{table}

\begin{figure*}[!htp]
\centering
\includegraphics[width=.48\textwidth]{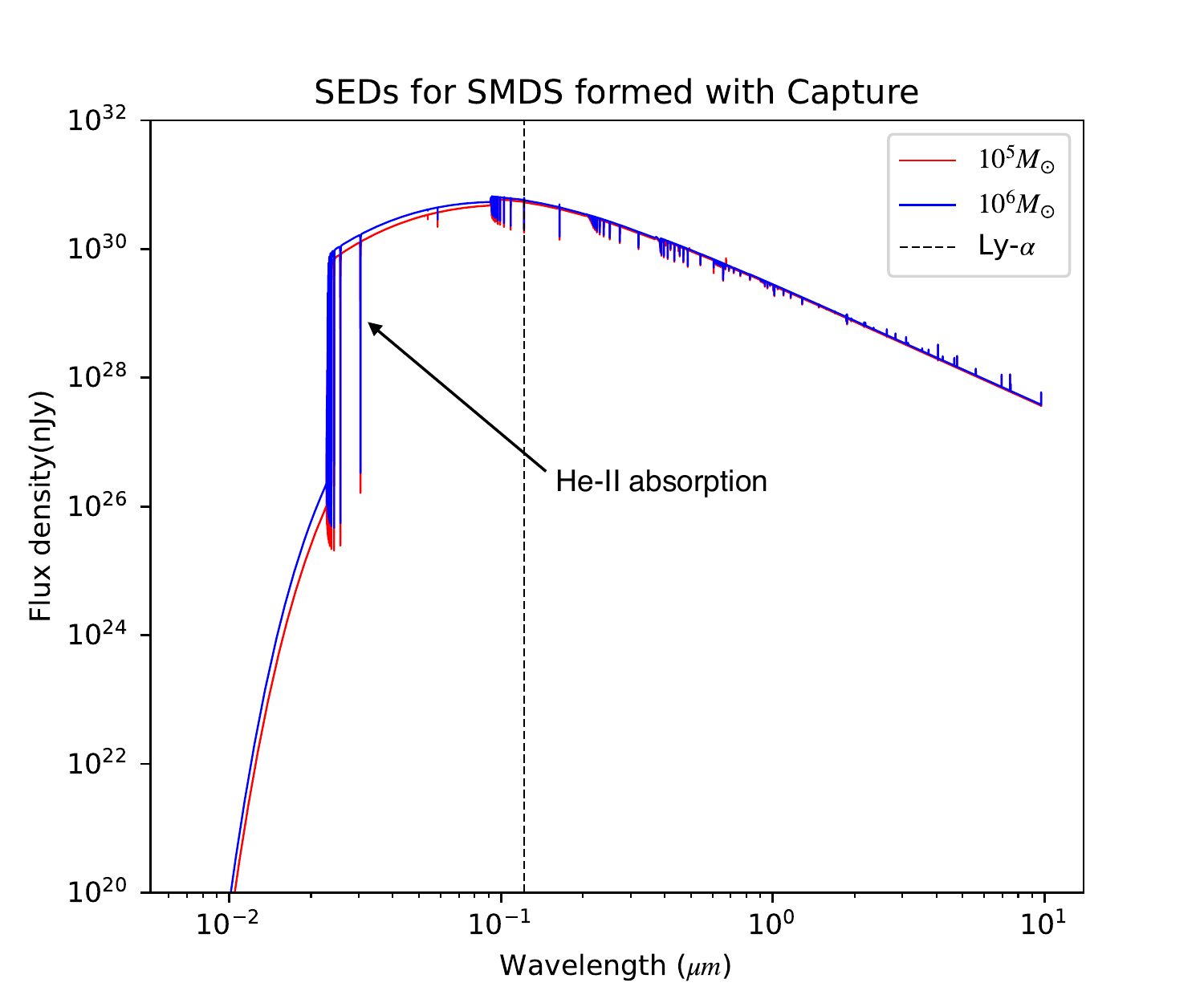}
\includegraphics[width=.48\textwidth]{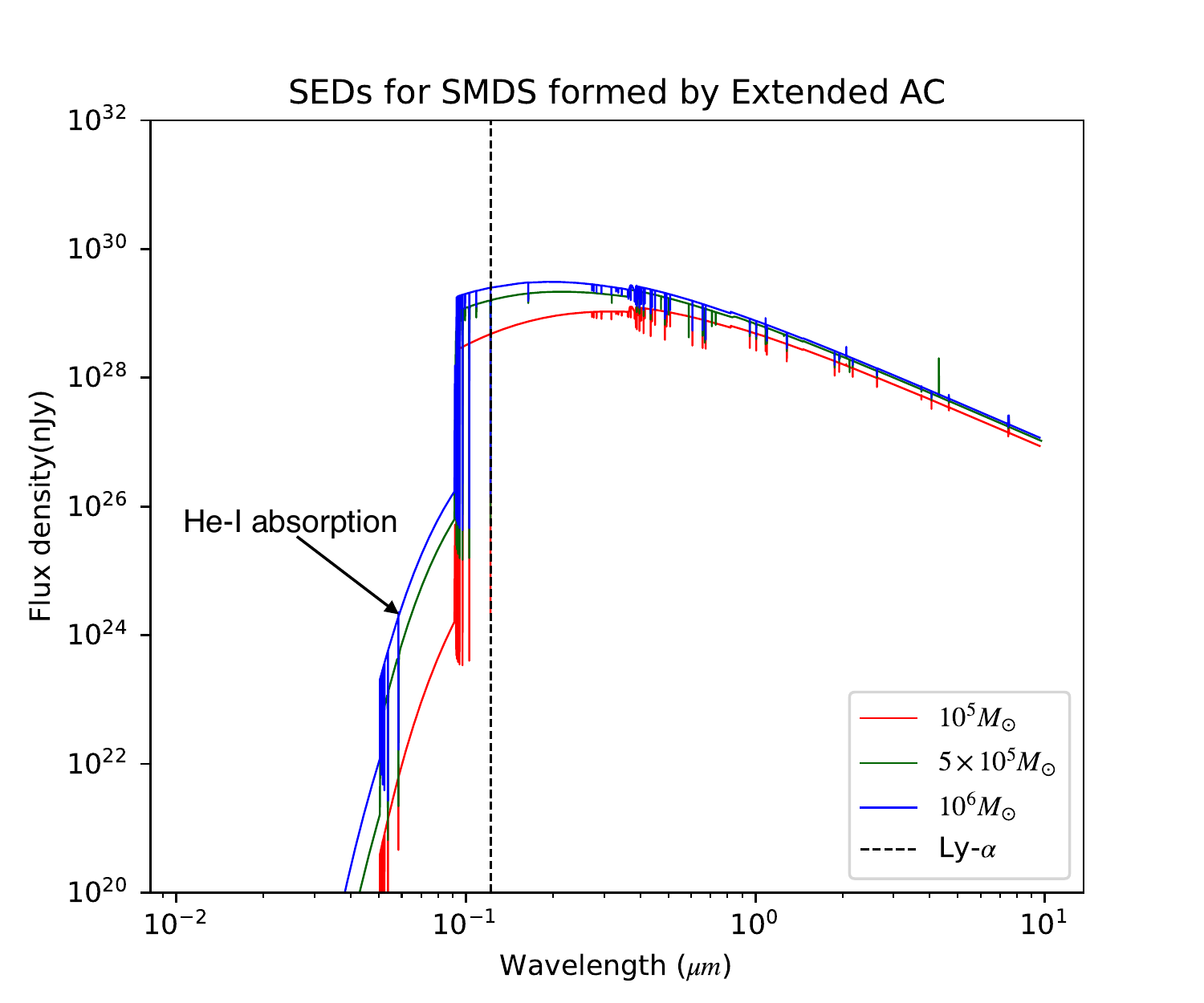}
\includegraphics[width=.48\textwidth]{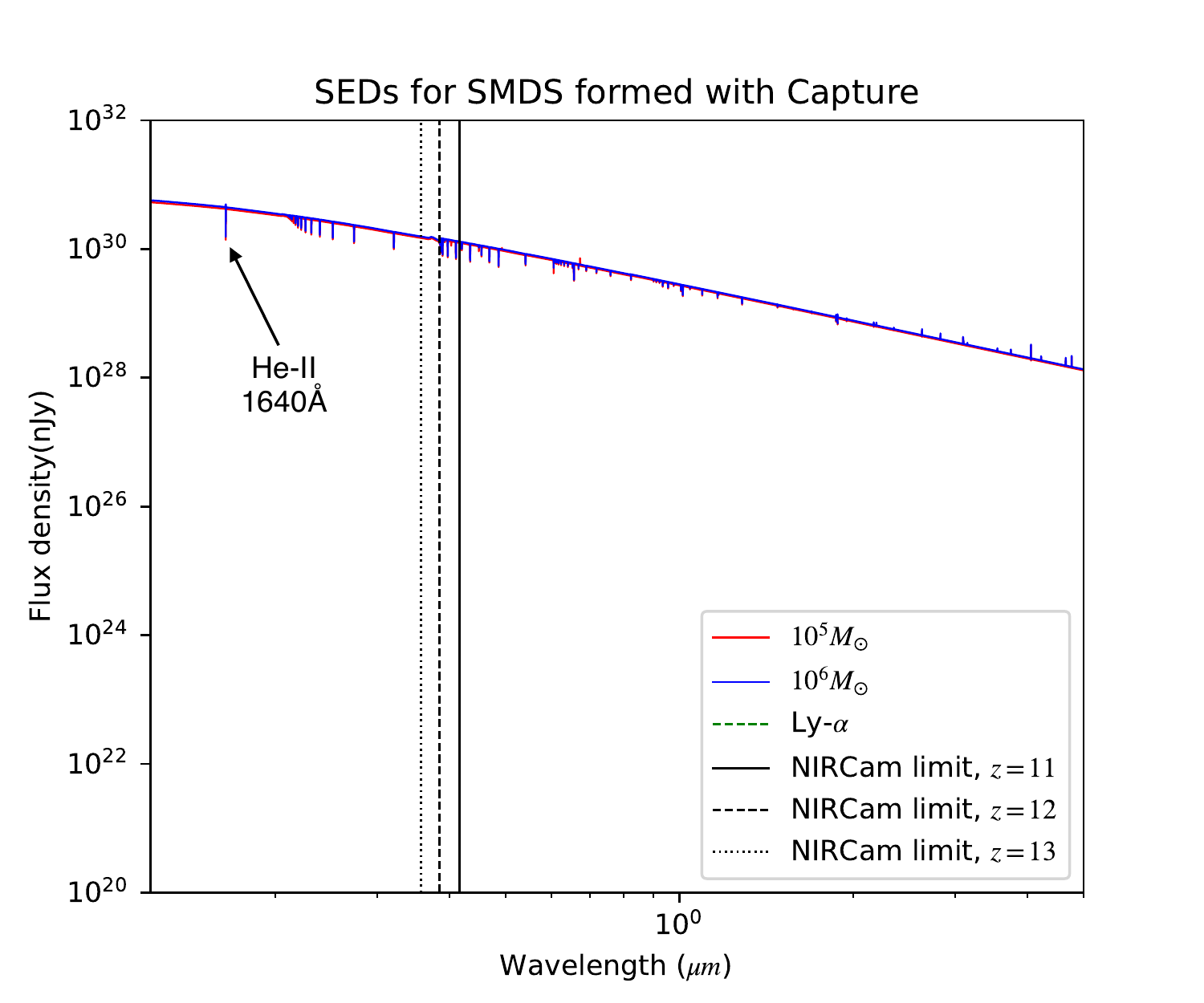}
\includegraphics[width=.48\textwidth]{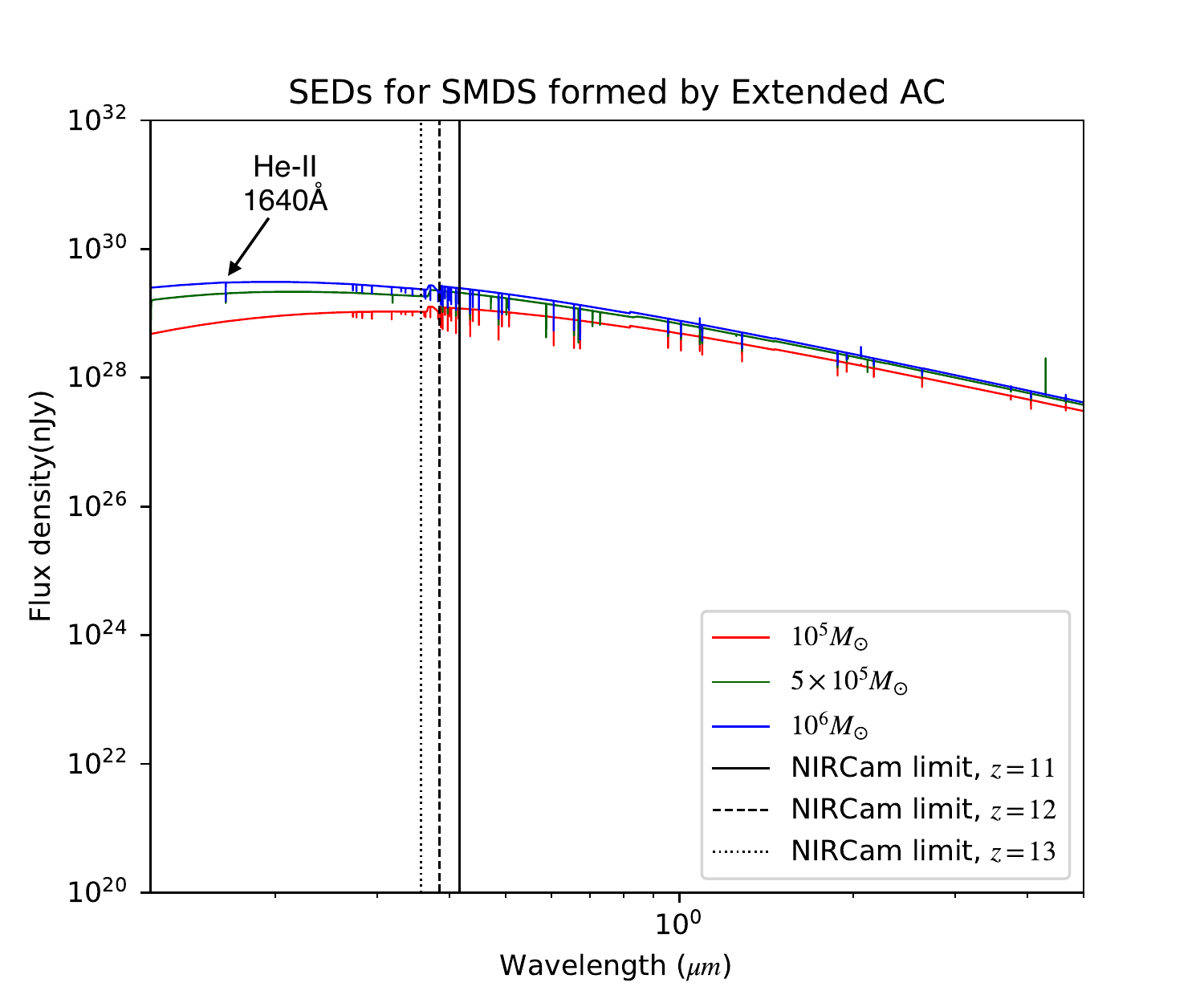}
\caption{\textsc{TLUSTY} simulated SEDs of supermassive dark stars of different masses. The left panels represent SMDS formed with capture and the right panels represent SMDS formed via extended AC. Parameters for these model SMDS can be found in table \ref{tab:SMDSpara}. In the lower panels, we zoom into the region observable by NIRCam and NIRSpec, between Ly-$\alpha$ and $\microm{5}$. The NIRCam limits (corresponding to $\lambda_{obs}=\microm{5}$) at different redshifts are shown in black. At the other end, we are limited by the Gunn-Peterson trough. Therefore, NIRCam/NIRSpec will only observe the very narrow part of the SED between \Lyalpha and the vertical black lines.}
\label{fig:spectra}
\end{figure*}

In Fig.~\ref{fig:spectra} we plot the restframe SEDs of SMDSs of various masses (with values labeled in the legend) formed via the two mechanisms described in the main text: Adiabatic Contraction and DM Capture. The parameters for each star are given in the caption, and were obtained in~\cite{Freese:2010smds}, when two of us used the polytropic approximation to model SMDSs. We note here that for radiation pressure dominated stars, such  as SMDSs, a polytrope of index $n=3$ is an excellent approximation.

\begin{figure}[!htb]
    \centering
    \includegraphics[width=.75\linewidth]{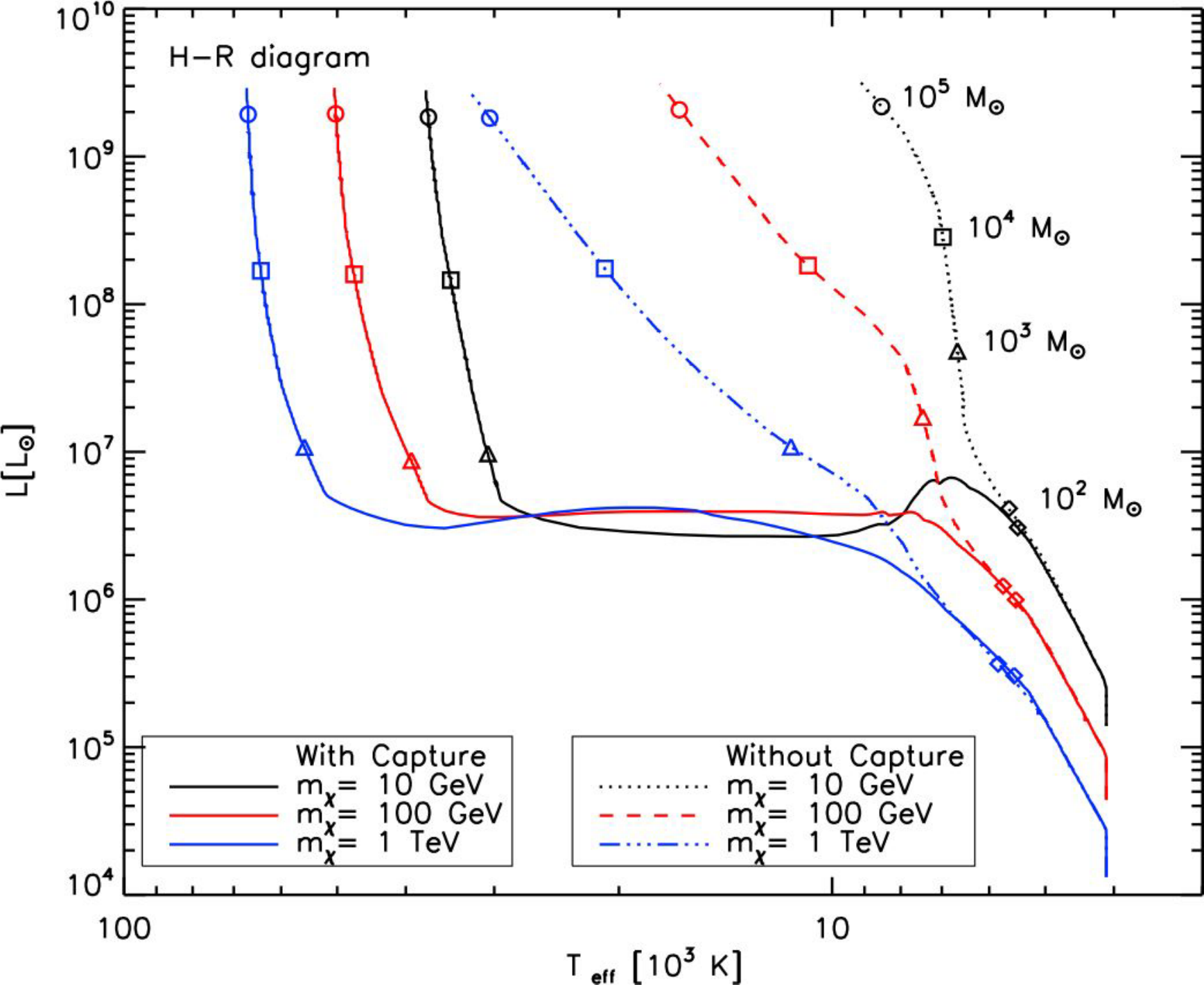}
    \caption{Hertzsprung-Russell (HR) diagram for dark stars for accretion rate
  $\dot M = 10^{-3} \Msun$/yr  and a variety of WIMP masses as
  labeled for the two cases: (i) ``without capture'' but with extended
  adiabatic contraction (dotted lines) and (ii) ``with capture''
  (solid lines).  The case with capture is for product of scattering
  cross section times ambient WIMP density $\sigma \bar\rho_\chi =
  10^{-39} {\rm cm}^2 \times 10^{13}$GeV/cm$^3$.  Also labeled are
  stellar masses reached by the DS on its way to becoming
  supermassive. The final DS mass was taken to be $1.5\times 10^5
  \Msun$ (the baryonic mass inside an assumed $10^6\Msun$ DM host halo), but it could be larger, depending on the mass of the host halo. (Figure reproduced from~\cite{Freese:2010smds}).}
  \label{fig:SMDSHR}
\end{figure}

We end this section with Fig.~\ref{fig:SMDSHR}, an HR diagram for Dark Stars (reproduced from~\cite{Freese:2010smds}).

\section{SMDS Candidates}

We start this section with Table~\ref{tab:BestFit}, where we list the best-fit parameters for each SMDS candidate. 

\begin{table}[h]
\centering
\begin{tabular}{lcccccccc}
\hline
Candidate & $z_{\text{phot}}$ & $z_{\text{spec}}$ & $\mu$ & $\chi^2$  &  $\chi_{\text{crit}}^2$ & $\chi_{\text{gal}}^2$ & Formation Mechanism & SMDS Mass ($M_{\odot}$)\\ [0.5ex] 
 \hline\hline
JADES-GS-z13-0 & 13.98 & 13.20 & 1.50 & 14.12 & 18.3 & 6.8 & Capture & $10^6$ \\
JADES-GS-z12-0 & 12.27 & 12.63 & 1.11 & 5.64 & 15.5 & 3.6 & Extended AC & $5\times10^5$\\
JADES-GS-z11-0 & 11.66 & 11.58 & 0.75 & 12.23 & 22.4 & 14.7 & Extended AC & $10^6$\\
\hline
\end{tabular}
\caption{The best-fit parameters corresponding to each of the SMDS candidates. $z_{\text{phot}}$ is the photometric redshift assuming the SMDS mass and formation mechanism listed in the last two columns of this table. $z_{\text{spec}}$ is the spectroscopic redshift found in \cite{JADES:2022a}. $\mu$ is the gravitational lensing factor for our SMDSs best fits. $\chi^2$ is the value found during our analysis; $\chi_{\text{crit}}^2$ is the value of $\chi^2$ required for 95\% confidence in our result; and $\chi_{\text{gal}}^2$ is the value of $\chi^2$ found assuming these objects are galaxies, as in \cite{JADES:2022a}.}\label{tab:BestFit}
\end{table}

\begin{figure*}[!htp]
\centering
\includegraphics[width=.48\textwidth]{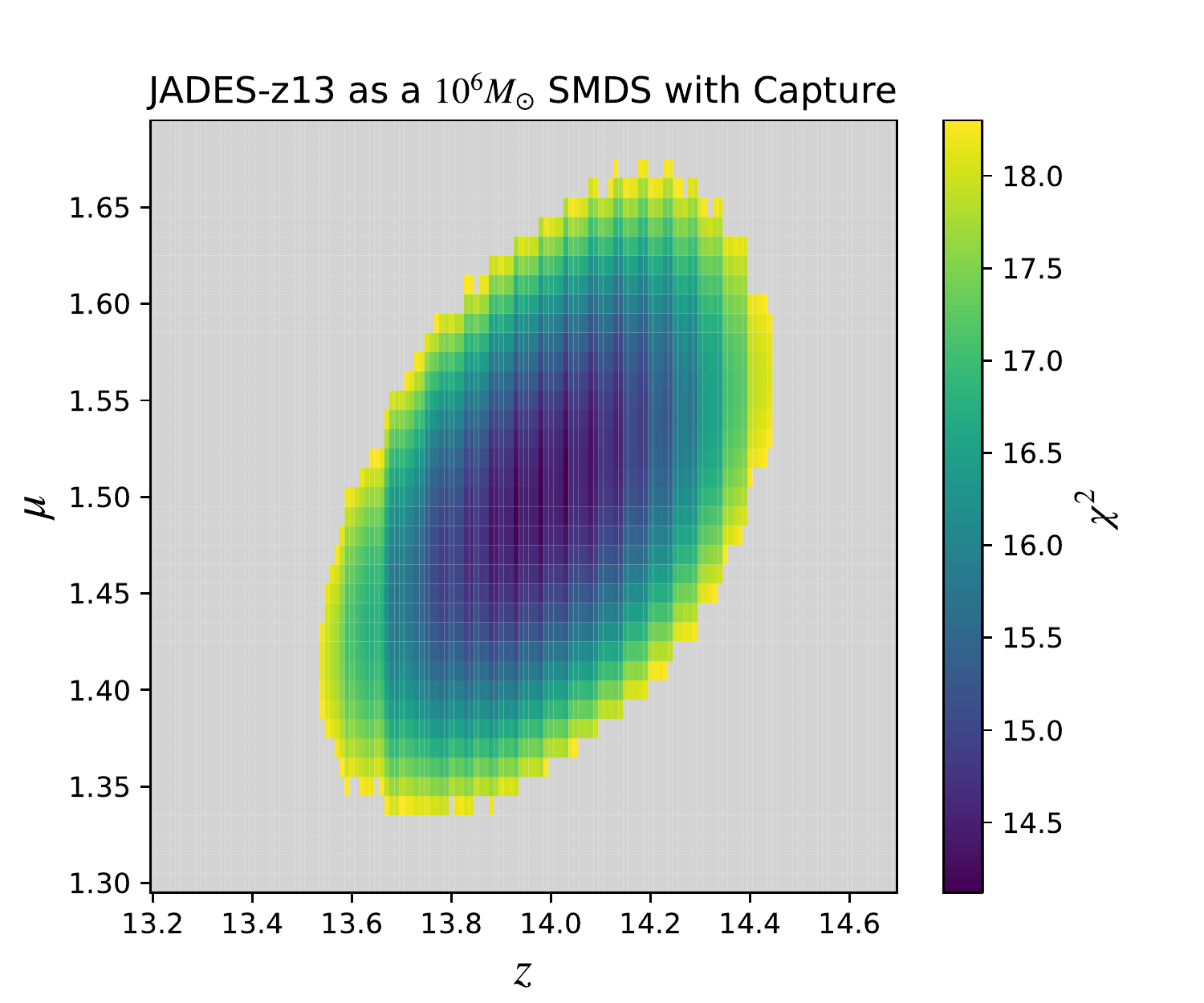}
\includegraphics[width=.48\textwidth]{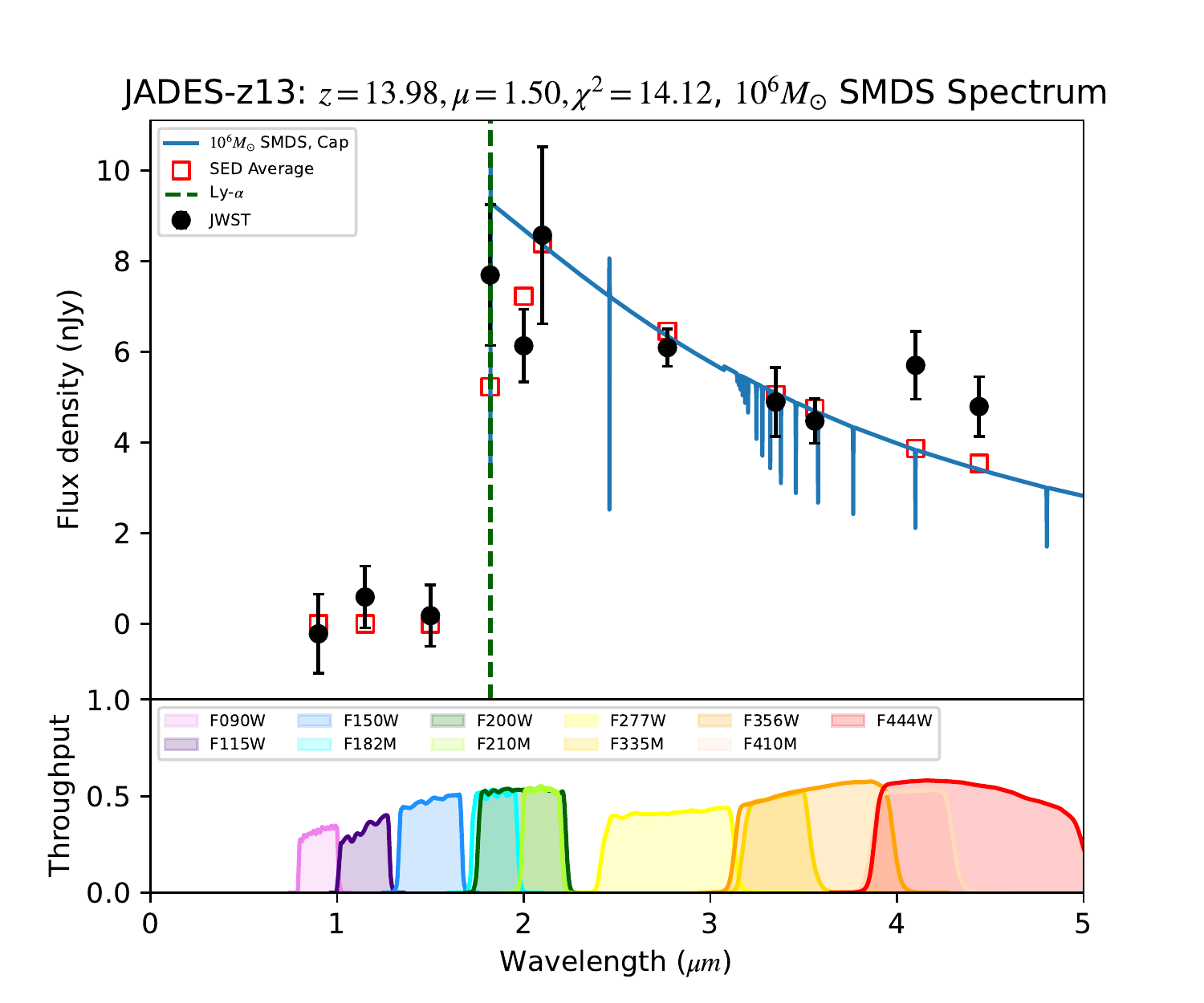}
\caption{ (Left Panel) Optimal fit region in the $z$ vs $\mu$ parameter space for \JADESzthirteen as a $10^6\Msun$ SMDS formed via DM capture. The heatmap is color coded according to the value of the $\chi^2$, and is cut off (grayed out) at the critical value corresponding to 95\% CL.  (Left panel) We plot our best fit SMDSs SEDs against the photometric data of~\cite{JADES:2022a} in each band (color coded and labeled in legend). For a $10^6\Msun$ SMDSs formed via DM capture, our best fit parameters take the following values: $z_{photo}=13.98$ and $\mu=1.50$. For our best fit we have $\chi^2=14.12$, whereas the critical value, corresponding to 95\% CL is, given the 11 bands in which we have data, $\chi_{crit}=18.3$.}
\label{fig:JADES13}
\end{figure*}

\begin{figure*}[!htp]
\centering
\includegraphics[width=.48\textwidth]{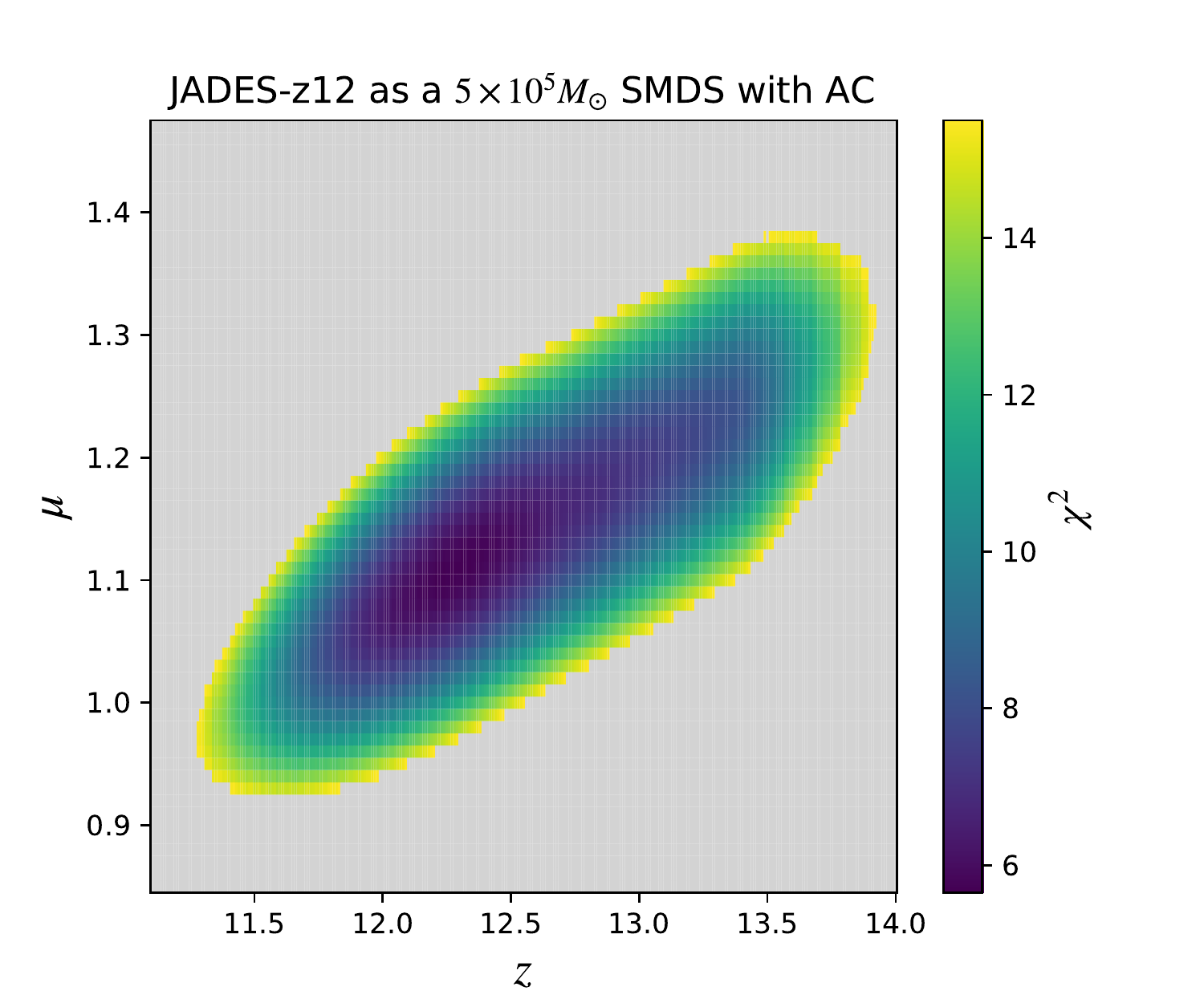}
\includegraphics[width=.48\textwidth]{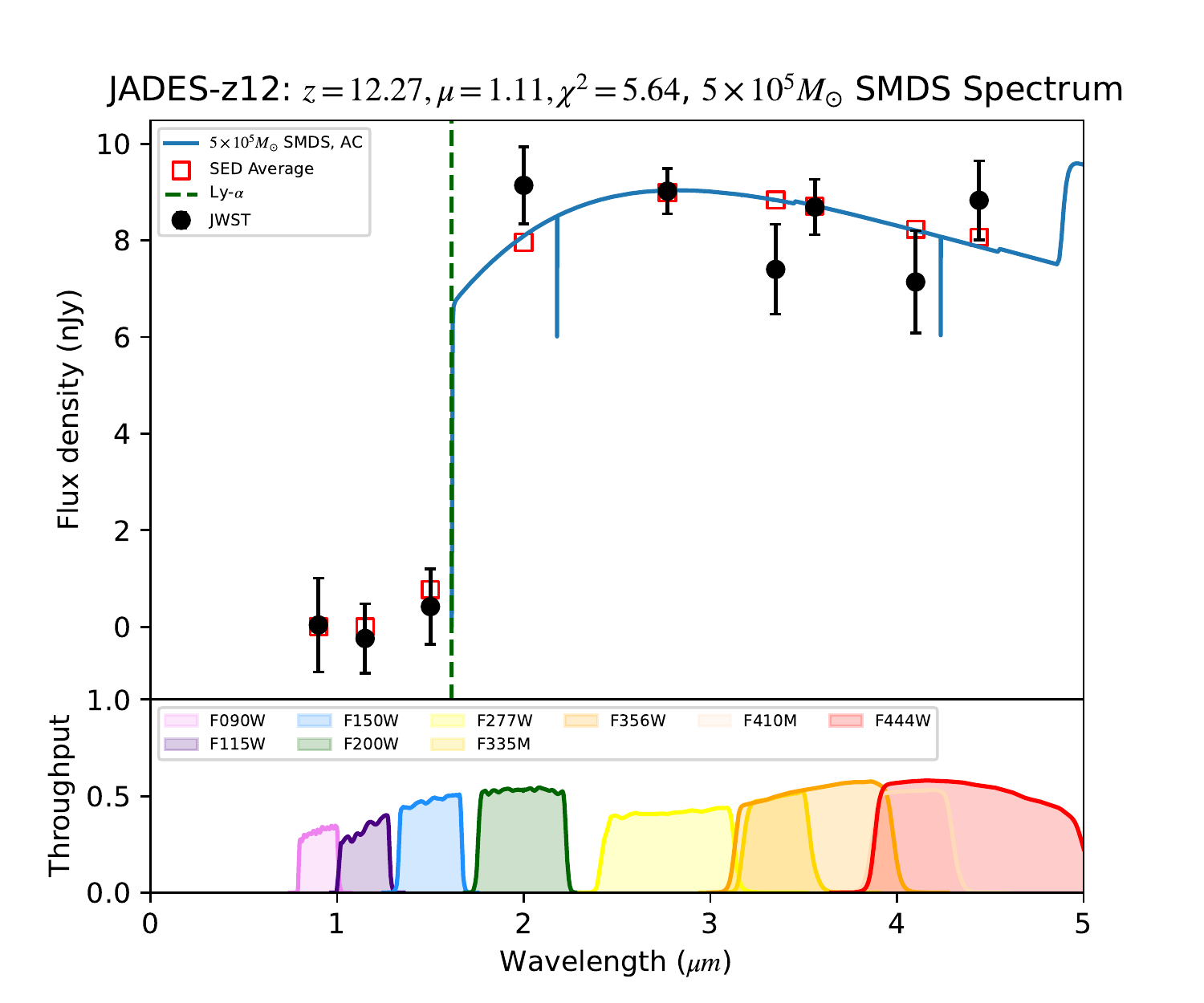}
\caption{ (Left Panel) Optimal fit region in the $z$ vs $\mu$ parameter space for \JADEStwelve as a $5\times 10^5\Msun$ SMDS formed via AC. The heatmap is color coded according to the value of the $\chi^2$, and is cut off (grayed out) at the critical value corresponding to 95\% CL.  (Left panel) We plot our best fit SMDSs SEDs against the photometric data of~\cite{JADES:2022a} in each band (color coded and labeled in legend). For a $5\times 10^5\Msun$ SMDSs formed via AC, our best fit parameters take the following values: $z_{photo}=12.27$ and $\mu=1.11$. For our best fit we have $\chi^2=5.64$, whereas the critical value, corresponding to 95\% CL is, given the 9 bands in which we have data, $\chi_{crit}=15.51$.}
\label{fig:JADES12}
\end{figure*}

\begin{figure*}[!htp]
\centering
\includegraphics[width=.48\textwidth]{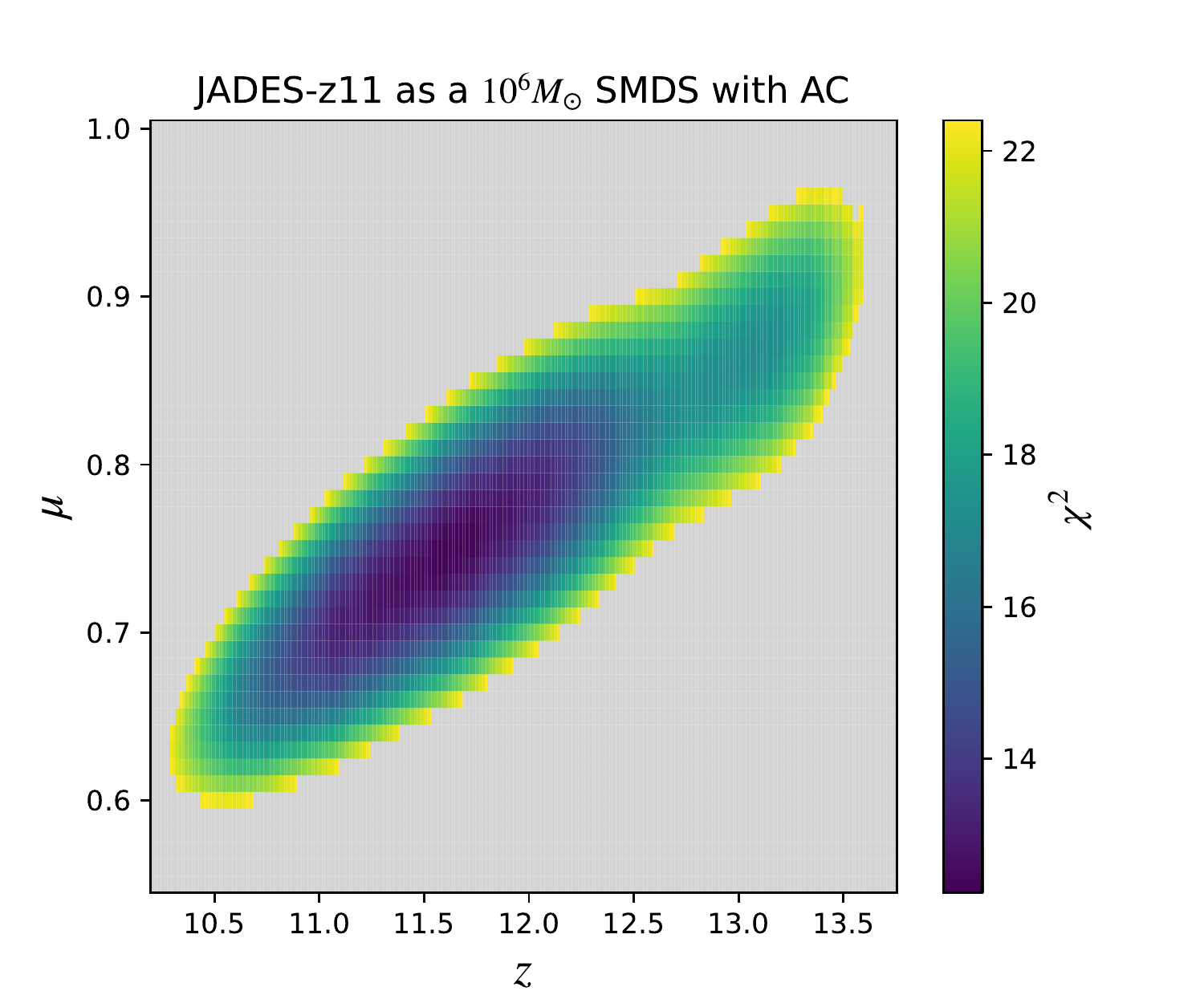}
\includegraphics[width=.48\textwidth]{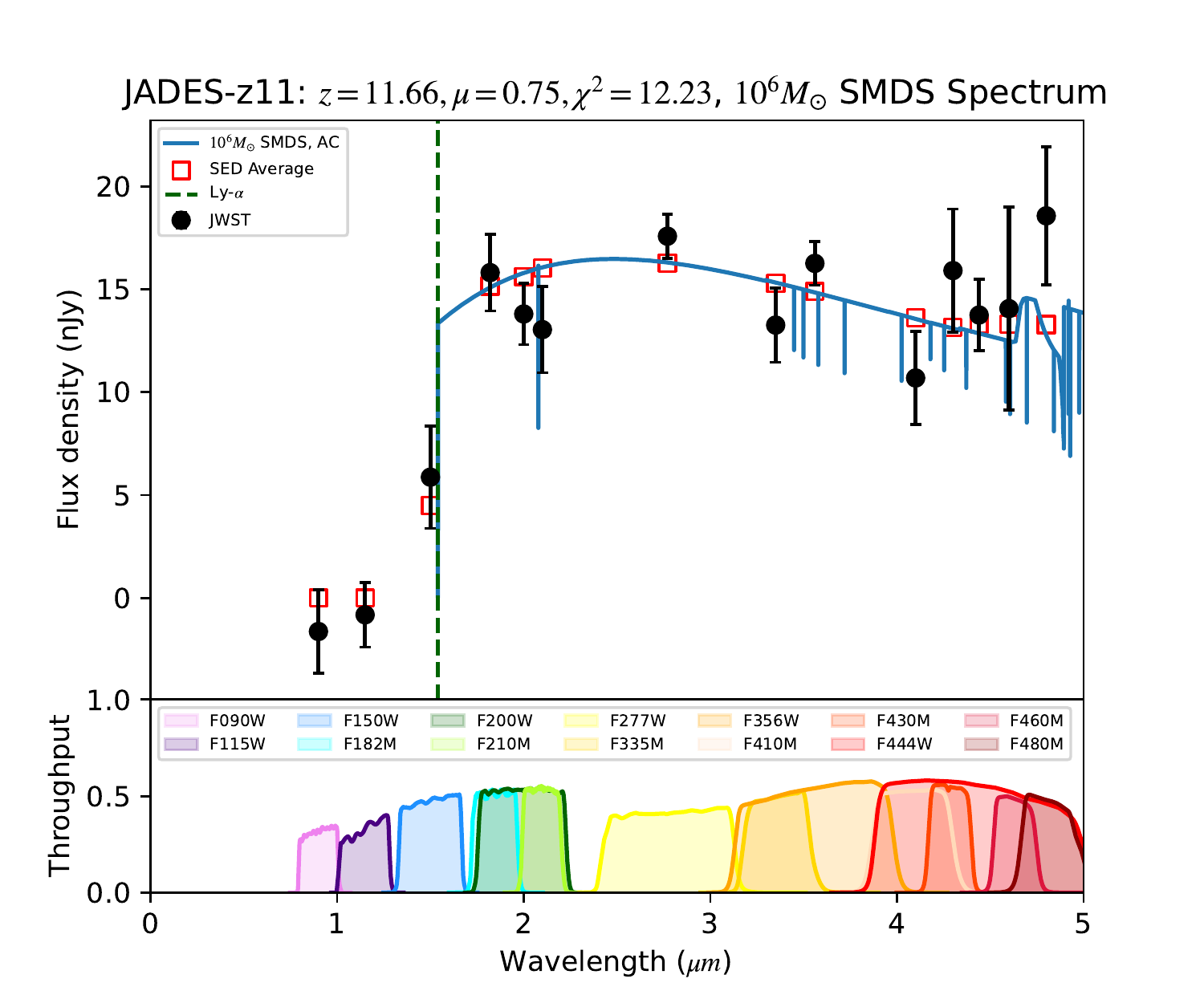}
\caption{ (Left Panel) Optimal fit region in the $z$ vs $\mu$ parameter space for \JADESeleven as a $10^6\Msun$ SMDS formed via AC. The heatmap is color coded according to the value of the $\chi^2$, and is cut off (grayed out) at the critical value corresponding to 95\% CL.  (Right panel) We plot our best fit SMDSs SEDs against the photometric data of~\cite{JADES:2022a} in each band (color coded and labeled in legend). For a $10^6\Msun$ SMDSs formed via AC, our best fit parameters take the following values: $z_{photo}=11.66$ and $\mu=0.75$. For our best fit we have $\chi^2=12.23$, whereas the critical value, corresponding to 95\% CL is, given the number of bands in which we have data, $\chi_{crit}=22.36$.}
\label{fig:JADES11}
\end{figure*}

In this section we additionally present enlarged versions of the plots included in Fig.~\ref{fig:Heatmaps} from the main text, as follows:\JADESzthirteen, in Fig.~\ref{fig:JADES13}, \JADEStwelve in Fig.~\ref{fig:JADES12}, and finally \JADESeleven in Fig.~\ref{fig:JADES11}.

\clearpage
\newpage
\bibliography{RefsDM.bib}

\end{document}